\documentclass[a4paper]{article}
\pdfoutput=1

\usepackage{amsfonts}
\usepackage{amssymb}
\usepackage{amsthm}
\usepackage[margin=2cm]{geometry}
\usepackage[dvips]{graphicx}
\usepackage{pgf}
\usepackage[T1]{fontenc}
\usepackage[latin2]{inputenc}
\usepackage{subfigure}
\usepackage{amsmath}
\usepackage{bbm}

\def\openone{\leavevmode\hbox{\small1\normalsize\kern-.33em1}}
\def\bra#1{\mathinner{\langle{#1}|}}
\def\ket#1{\mathinner{|{#1}\rangle}}
\def\braket#1#2{\mathinner{\langle{#1}|{#2}\rangle}}

{\catcode`\|=\active
  \gdef\Braket#1{\left<\mathcode`\|"8000\let|\BraVert {#1}\right>}}
\def\BraVert{\egroup\,\mid@vertical\,\bgroup}
{\catcode`\|=\active
  \gdef\set#1{\mathinner{\lbrace\,{\mathcode`\|"8000\let|\midvert #1}\,\rbrace}}
  \gdef\Set#1{\left\{\:{\mathcode`\|"8000\let|\SetVert #1}\:\right\}}}
\def\midvert{\egroup\mid\bgroup}
\def\SetVert{\egroup\;\mid@vertical\;\bgroup}
\newcommand{\proj}[1]{\ket{#1}\bra{#1}}

\newcommand{\Tr}{{\rm Tr}}
\newtheorem{theorem}{Theorem}

\begin{document}


\title{Quantumness Witnesses}

\author{Robert Alicki$^1$, Marco Piani$^2$ and Nicholas Van Ryn$^3$, \\
\\
{\small
$^1$ Institute of Theoretical Physics and Astrophysics, University of
Gda\'nsk,  Wita Stwosza 57, PL 80-952 Gda\'nsk, Poland.
}\\
\\
{\small
$^2$ Institute for Quantum Computing and Department of Physics and Astronomy, University of Waterloo,}\\
{\small  200 University Ave. W., N2L 3G1 Waterloo ON, Canada
}\\
\\
{\small
$^3$ School Of Physics, Quantum Research Group, University of
KwaZulu-Natal,}\\{\small Westville Campus, Private Bag x54001,
Durban, South Africa.} }

\date{\today}
\maketitle

\begin{abstract}
A recently proposed test of quantumness \cite{AV} is put into a broader mathematical and physical perspective. The notion of quantumness witness is introduced, in analogy to entanglement witness,
and illustrated by examples of a  single qubit and  many-body systems with additive observables. We compare also our proposal with the quantumness test based on quantum correlations (entanglement) and  Bell inequalities. A class of quantumness witnesses
associated to the phase space representation is also discussed.
\end{abstract}

\section{ Introduction}

One ``rule of thumb'' of quantum theory is the Bohr Correspondence Principle,
which can be formulated as follows:
\begin{quote}
\emph{The systems consisting of a large number
of particles and/or emerging in quantum states characterized by
large quantum numbers  behave classically.}
\end{quote}
However, the validity of this principle, in particular the range of its application,
remains open. It seems that with improved experimental techniques the actual
border between quantum and classical worlds has been moved into larger and larger systems
\cite{dec}. The most daring challenge to the correspondence principle is the idea of macroscopic quantum systems. For example, there exists quite convincing evidence that certain macroscopic systems, like Josephson junctions ("superconducting qubits")\cite{NBS}, Bose-Einstein condensates \cite{BEC} or Rydberg atoms \cite{Ahn}, preserve fundamental quantum properties. On the other hand, these arguments are strongly model-dependent and do not completely exclude the existence of an approximative classical description  \cite{Gro,ali}. Therefore,  model-independent tests of "quantumness" are of great theoretical and practical importance. This question is particularly relevant for the field of quantum information and quantum computation. A useful quantum computer should be a rather macroscopic machine which nevertheless preserves certain fundamental quantum properties. Moreover, it is believed that some examples of macroscopic quantum systems can provide promising implementations of quantum information processing.
\par
In the present paper we develop the ideas introduced in \cite{AV} and tested experimentally for a single-photon polarization in \cite{Gen}. We discuss tests of quantumness based on certain fundamental properties of classical probability theory which are not valid in quantum theory.
Mathematically, these tests involve certain specific quantum observables called \emph{quantumness witnesses} in analogy to \emph{entanglement witnesses} \cite{Alb}. We discuss in detail the case of a single qubit and then show how  the quantumness disappears for many-body systems when we restrict ourselves to additive observables only. We compare our proposal with the quantumness test related to quantum correlations (entanglement) and  Bell inequalities. We also study quantumness witnesses which appear in  the context of phase space representation of a quantum oscillator.

\section{States and observables}

Assume that our aim is to interpret a given set of experimental data in terms of a mathematical model which involves the notions of states and observables. Operationally, the state $\rho$ can be identified with a fixed system's preparation procedure and the observable $A$ with some apparatus which produces a set of outcomes  $\{a\}\equiv \{a_1,a_2,...\}$. Repeating the given preparation procedure $\rho$ and the measurement of $A$ many times we obtain the probability distribution  $p_a=(p_1, p_2,...)$ which allows, e.g., to compute all moments of the observable
\begin{equation}
\langle A^k\rangle_{\rho} = \sum_j p_j a_j^k\ ,\  k= 0,1,2,....
\label{moment}
\end{equation}
Therefore, a single apparatus corresponds not to a single observable but rather to the whole family of functions $F(A)$ of $A$ which differ by a choice of a "pointer scale" only.

\subsection{ $C^*$-algebraic model}

We now need a mathematical model of states and observables. For simplicity we denote by the same symbols the physical states and observables and their mathematical representations. Practically, we shall test only two types of models, classical and quantum. Both can be unified within the mathematical scheme called the \emph{$C^*$-algebraic model}. However, for all practical purposes one can always think about the two simplest extreme cases - the classical algebra of complex functions $\mathcal{C}(\Gamma)$ on a "discrete phase-space" $\Gamma =\{1,2,...,n\}$ and the quantum algebra $\mathcal{M}_n$ of $n\times n$ complex matrices. A $C^*$-algebra $\mathcal{A}$ is a complete complex linear space with a norm $\|\cdot\|$,
adjoint operation $A\to A^{\dagger}$ and the product $AB$ satisfying natural relations including the condition $\|AA^{\dagger}\| =\|A\|^2$. We always assume that $\mathcal{A}$ contains a unit element $I$. The main difference between the algebra of functions and the algebra of matrices is that the former is commutative (i.e. AB=BA) while the latter is not. Linear and bounded functionals on $\mathcal{A}$ form a linear and normed space. The linear functional $\omega$ is positive
if $\omega (A A^{\dagger})\geq 0$ for all $A\in\mathcal{A}$ and normalized if $\omega(I)=1$. The elements of the form $AA^{\dagger}$ are called positive and can be used to define a partial order relation in $C^*$-algebra $\mathcal{A}$
\begin{equation}
A\leq B\ \mathrm{if\ and\ only\ if\ there\ exists }\ C\  \mathrm{such\ that}\ B= A +CC^{\dagger}.
\label{order}
\end{equation}
\par
Within the $C^*$-algebraic model we identify all bounded observables with self-adjoint (i.e. $A=A^{\dagger}$) elements in such a way that \emph{the $k$-moment of the observable $A$ as defined by (\ref{moment}) is identified with the algebraic $k$-power  of the corresponding element of the $C^*$-algebra $\mathcal{A}$}. The set of all states is identified with the set of all positive linear functionals on $C^*$-algebra $\mathcal{A}$ . This is a convex set and its extremal points are identified with pure states. For the given two examples we obtain the standard pictures. In the classical theory (positive) observables are
(positive) real functions $A(\gamma), \gamma = 1,2,..n$ and the states form a simplex of probability distributions  $P=(p(1), p(2),..,p(n))$ with extreme points (pure states) of the form $p(j) = \delta_{jk}$. The possible outcomes of the measurement of $A$ are given by the numbers $\{A(\gamma)\}$ and for the moments we have $\langle A^k\rangle_{P} = \sum_{\gamma} p(\gamma) A(\gamma)^k$. In the quantum theory (positive) observables are
(positive) hermitian matrices with spectral representations $A =\sum_j a_j P_j\ ,\ P_j^2= P_j , \sum_j P_j = I$, and the states are identified with density matrices $\rho= \sum_m \rho_m |m\rangle\langle m| ,\ \langle m'|m\rangle=\delta_{m'm}$. Extreme points (pure states) are one-dimensional projections $|\psi\rangle\langle\psi|$ identified with the normalised vectors in the Hilbert space $\mathbf{C}^n$. The possible outcomes of the measurement of $A$ are given by its eigenvalues $\{a_j\}$ and for the moments we have $\langle A^k\rangle_{\rho} = \mathrm{Tr}(\rho A^k)=\sum_{jm}\rho_m (a_j)^k \langle m|P_j|m\rangle$.

\subsection{ The main theorem}

For a concrete physical system it is very easy to find the differences between the predictions of the classical and quantum  model. However, \emph{our aim is to find the quantumness tests which are  model-independent but still operational and refer to the most fundamental mathematical differences between classical and quantum theory}. Assuming that we always work in the framework of $C^*$-algebraic scheme we can use the following general theorem which summarizes some basic results in the theory of $C^*$-algebras \cite{Dix,Kad}.
\begin{theorem}
The following statements are equivalent:
\begin{itemize}
\item[(a)] For any pair $A, B\in\mathcal{A}$ , $0\leq A\leq B$  implies $A^2\leq B^2$;
\item[(b)] For any pair $X, Y\in\mathcal{A}$ , $X\geq 0, Y\geq 0$  implies $XY + YX \geq 0$;
\item[(c)] $C^*$-algebra $\mathcal{A}$ is commutative;
\item[(d)] $C^*$-algebra $\mathcal{A}$ is isomorphic to the algebra of continuous functions on a certain compact set.
\end{itemize}
\end{theorem}
\begin{proof}
The equivalence of (c) and (d) is an important result proved, for example, in \cite{Dix,Kad}. The equivalence of
(a) and (c) has been proved in \cite{Oga}. For practical purposes, in order to prove the equivalence of (a) and (c) we can use as a proof the qubit example from Section 3 because all physically relevant quantum $C^*$-algebras contain a two dimensional matrix algebra $\mathcal{M}_2$.\\
In order to prove a)$\Rightarrow$ b), take $A= X$ and $B = X +t Y$. Then
\begin{equation}
0\leq B^2 - A^2 = t (tY^2 + XY + YX)\ \mathrm{for\ any}\ t\ \geq 0
\label{proof}
\end{equation}
which implies $XY + YX \geq 0$.\\
In order to prove (b)$\Rightarrow$(a) use the inequality $0\leq A\leq B$ for
\begin{equation}
2(B^2 - A^2) = (B-A)(B+A) + (B+A)(B-A)\ .
\label{proof1}
\end{equation}
\end{proof}

{\bf Remarks} In the following, Theorem 1 will be used to design the "quantumness tests" which can exclude classical algebraic models for  given sets of experimental data. In the condition a), instead of the square function one can use any  \emph{operator non-monotone function}\footnote{A function $f:\mathbb{R}\rightarrow \mathbb{R}$ is \emph{operator monotone} if for all $n$ and all $A,B\in\mathcal{M}_n$ we have that $A\leq B$ implies $f(A)\leq f(B)$.}, for example  $A \to A^{\alpha}$, $\alpha > 1$.

\subsection{Hidden variable models}
Although our aim is not to challenge the quantum mechanical model of  Nature but rather to discuss the means of distinguishing quantum systems operating in a semiclassical regime from those which still preserve some experimentally accessible and practically useful quantum features, a brief discussion of the relations between our notion of a \emph{Classical Algebraic Model} (CAM) and the notion of a \emph{Hidden Variable Model} (HVM) seems to be unavoidable.\\
There exist different definitions of HVMs which are probably  not equivalent from the mathematical point of view (see the monograph \cite{Dick} and references therein). Their common feature is the existence of a set of parameters $\{\lambda\}$ (\emph{hidden variables}) which determine  all possible outcomes of experiments performed on a given system. This idea can be realised assuming that the experiments $A, B, C,..$ produce outcomes $\{a\}, \{b\}, \{c\},...$ which are  functions of the hidden variables $a(\lambda),b(\lambda), c(\lambda),...$. The perfect system's preparation procedure can be described by fixing the hidden variable $\lambda$ which in principle may depend not only on the preparation device but also on the \emph{context}, i.e. the set of observables to be measured. As we would like to model nondeterministic theories we assume that the preparation procedure is not perfect but rather given by the probability distribution on the set of
hidden variables which may still depend on the context and is denoted by  $P(\lambda ; A,B,C,...)$. The mean value of the observable $A$ and the correlations of $A,B$ are given by the following expressions
\begin{equation}
\langle A\rangle_P = \int P(\lambda ; A,B,C,...)a(\lambda)\, d\lambda\ ,\ \langle AB\rangle_P = \int P(\lambda ; A,B,C,...)a(\lambda)b(\lambda)\, d\lambda\ .
\label{HV}
\end{equation}
It seems that the presented scheme  covers most of the known examples of HVMs, including  the so-called \emph{nonlocal} and/or \emph{contextual} ones. The CAM considered in this paper can be treated as a special case of the HVM of above with the additional assumption that the state preparation procedure and the different measurement procedures are independent of each other which implies the following \emph{independence condition}~\cite{Ind}
\begin{equation}
P(\lambda ; A,B,C,...)\equiv P(\lambda )\ .
\label{ind}
\end{equation}
The condition (\ref{ind}) is often attributed to \emph{locality} in the sense of special relativity theory. On the other hand, it is rather related to the possibility of decomposing the Universe into weakly interacting subsystems which correspond, for example, to a given physical system, preparing apparatus and measurement devices.

\section{Quantumness witnesses}

From Theorem 1 it follows that for a \emph{Quantum Algebraic Model} (QAM) we can always find pairs of observables $\{A, B ;0\leq A\leq B\}$
or $\{X, Y ; X\geq 0, Y\geq 0\}$ such that the operators  $V = B^2 - A^2$ and $C = XY + YX$ possess negative eigenvalues.
Analogically to the theory of entanglement \cite{H4} we can call such $V$ and $C$ \emph{quantumness witnesses}(QW). We remind the reader that an \emph{entanglement witness} is an observable which possesses at least one negative eigenvalue but for all separable states yields  positive mean values.  The main problem is to design experimental tests of the "violation of classicality", i.e., an experimental proof of the existence of negative outcomes for $V$ and $C$. Within the algebraic framework one assumes that for any self-adjoint element of $\mathcal{A}$ there exists a physically realizable observable. In particular, for two observables $X$ and $Y$ the observables $X\pm Y$ and $XY + YX$
exist also. However, in general, there is no operational prescription as to how to construct the corresponding apparatus  if we know how to do it for $X$ and $Y$. Fortunately, for the quantumness witness $V$ it is enough to measure the second moments of $A$ and $B$ and find a state $\rho$ such that $\langle B^2\rangle_{\rho} < \langle A^2\rangle_{\rho}$. For $C$ the situation is different, namely in the general case there is no operational prescription as to how to measure the  (symmetrized) correlations
$\langle XY + YX\rangle_{\rho}$. The exception is the case of jointly measurable (i.e. commuting) observables where we simply take as outcomes for $XY=YX$ the products of the outcomes for $X$ and $Y$. Another situation is discussed in Section 5, in the context of Bell inequalities, where the mean value of $C$ is related to the mean value of the Bell observable, whose mean value can be computed from the experimental data.\\

\subsection{The abundance of quantumness witnesses}

Theorem 1 ensures only the existence of a QW for the quantum algebraic model. The following Theorem 2 shows their abundance.
\begin{theorem}
For any $n\times n$ density matrix $\rho \neq \openone/n$ there exist quantumness witnesses of the type $V= B^2 - A^2$ and $C= XY + YX$. For the maximally mixed state such witnesses do not exist.
\end{theorem}
\begin{proof}
We first prove the case $n=2$. The density matrix can be written as
\begin{equation}
\rho= (1-r)\frac{\openone}{2} + r\proj{\psi},\label{eq:rho}
\end{equation}
with $r\geq0$ and $r=0$ for the maximally mixed state. We look for two observables of the form
$X=\proj{a}, Y=\proj{d}$, for some normalized vectors $\ket{a},\ket{d}$. We consider the expansion
\[
\ket{d}=\alpha\ket{a}+\beta\ket{a^\bot}, \quad \braket{a}{a^\bot}=0,\quad |\alpha|^2+|\beta|^2=1.
\]
Thus,
\[
XY + YX\equiv \{X,Y\}=2|\alpha|^2 \proj{a}+ \alpha\beta^*\ket{a}\bra{a^\bot}+\alpha^*\beta\ket{a^\bot}\bra{a}.
\]
The eigenvalues of $\{X,Y\}$ are $\lambda_\pm=|\alpha|(|\alpha|\pm 1)$, and we can always choose $\ket{a}$ and $\ket{d}$  so that $\ket{\psi}$ coincides with the eigenvector of  $\{X,Y\}$ corresponding to $\lambda_-$.

Therefore we have that for such choices
\[
\Tr(\rho \{X,Y\})= \frac{1-r}{2}\Tr\{X,Y\} + r\langle\psi,\{X,Y\}\psi\rangle= (1-r)|\alpha|^2+r|\alpha|(|\alpha|- 1)=|\alpha|(|\alpha|-r),
\]
which becomes strictly negative for
\[
0<|\alpha|< r.\] and provides the QW  $C= \{X,Y\}$. To obtain the QW  $V= B^2 - A^2$ we follow the arguments in the proof of the Theorem 1 with $A=X$ and $B= X+tY$.  It is worth noticing that for the choice $A=\proj{a}, B=\proj{a}+t\proj{d}$, one sees immediately that $B^2 - A^2\ngeq 0$ for all $t>0$, if $|\alpha|\neq 1$.

In order to generalize the result to an arbitrary $n$ one should notice that the condition $\rho \neq \frac{\openone}{n}$ means that there are always
two eigenvalues of $\rho$ , say $\rho_1, \rho_2$ satisfying $\rho_1 > \rho_2\geq 0$ with corresponding eigenvectors $|1\rangle,|2\rangle$. Therefore, we can repeat the proof of above with operators $X, Y$ having the supports on the two dimensional subspace spanned by $|1\rangle,|2\rangle$. The second statement of the Theorem follows from the fact that $\Tr(XY)\geq 0$ if both $X\geq 0$ and $Y\geq 0$.
\end{proof}

\section{Test of quantumness for a single system}
According to the discussion in the previous section, we choose the $C^*$-algebraic model  as a proper mathematical  idealization. Our aim is to propose tests which can eliminate the CAM as a valid description of the experimental data. We assume that the experimental situation can be described in terms of the set $\mathcal{S}_{exp}$ of accessible initial states of a certain physical system and the set of accessible  measurements (observables) $\mathcal{A}_{exp}$. For any observable $A\in \mathcal{A}_{exp}$ and any state $\rho\in\mathcal{S}_{exp}$ we can extract (by repeating measurements on the fixed initial state $\rho$) the statistics of the measurement outcomes. Therefore, if $A\in \mathcal{A}_{exp}$ then for any  real function $F$, $F(A)\in \mathcal{A}_{exp}$. We say that the pair $(\mathcal{A},\mathcal{S})$, where $\mathcal{A}$ is a $C^*$-algebra and $\mathcal{S}$ is a set of linear, positive and normalized functionals on $\mathcal{A}$, is a \emph{minimal algebraic model} for our set of experimental data if:
\begin{enumerate}
\item we can identify ${\cal A}_{exp}$ with a subset of  ${\cal A}$ and ${\cal S}_{exp}$
with a subset of ${\cal S}$, such that the corresponding mean values reproduce experimental data,
\item for any pair of observables
$A, B\in {\cal A}_{exp}$ ,  $\langle A\rangle_{\rho} \leq \langle B\rangle_{\rho}$  for arbitrary  $\rho \in {\cal S}_{exp}$
implies  $\langle A\rangle_{\sigma} \leq \langle B\rangle_{\sigma}$ for arbitrary $\sigma \in {\cal S}$.
\end{enumerate}

\noindent {\bf Example 1} Suppose $\mathcal{S}_{exp}=\{\proj{i}\}$, with $\{\ket{i}\}$ a complete orthonomal basis for $\mathbb{C}^n$, and $\mathcal{A}_{exp}$ is given by any set of observables diagonal in said basis. Then the set of observables diagonal in the same basis plus any set of states containing $\mathcal{S}_{exp}$, is a minimal algebraic model.

According to the definition above, if we can find two accessible observables $A$ and $B$ such that for all accessible states
$0\leq \langle A\rangle \leq \langle B\rangle$, and if we can prepare a certain state $\rho$ satisfying $\langle A^2\rangle_{\rho} > \langle B^2\rangle_{\rho}$, then, by definition, we can say that the set of experimental data does not admit a minimal classical model. Therefore, one is led to conclude that the corresponding physical system preserves some genuine quantum features, as any minimal model would have to be quantum according to Theorem 1. However, one could still argue that there may exist a "non-minimal" classical model, with the observable $B-A$ possessing negative outcomes which are averaged out by too coarse-grained initial states (probability distributions). The possibility of this alternative might be called the \emph{minimality loophole}. In any case, notice that this latter possibility would mean that the resolution on the side of observables is (much) higher than the corresponding resolution on the side of initial states, and this does not seem a reasonable condition. Indeed, in practice we typically use the same techniques for both the state preparation and the measurements. Therefore, within the given technological means, both resolutions should be of the same order.

\noindent {\bf Example 2}
Given the interval $[0,3]$ on the real line, consider the functions
\begin{equation}
A(x)=\begin{cases}
      \frac{3}{2} & x\in [0,2]\\
	0	  & x\in\; ]2,3]
     \end{cases},
\quad
B(x)=\begin{cases}
      3 & x\in [0,1]\\
	1	  & x\in ]1,3]
     \end{cases}.
\end{equation}
Further, consider the algebra $\mathcal{A}$ of all real functions with support on the interval $[0,3]$ and of the form $$f(x)=\sum_{i=1}^3 f_i \chi_i(x),$$ with $\chi_i$ the indicator function
\[
\chi_i(x)=\begin{cases}
      1 & x\in [i-1,i]\\
	0	& \textrm{elsewhere}
     \end{cases},
\]
so that each function $f$ in $\mathcal{A}$ is identified with a vector $(f_i)_i\equiv(f_1,f_2,f_3)$.
The algebraic structure is given by the operations $\alpha (f_i)_i \equiv (\alpha f_i)_i$, $(f_i)_i + (g_i)_i\equiv (f_i+g_i)_i$, and $(f_i)_i\times (g_i)_i \equiv (f_ig_i)_i$. It is immediately seen that $\mathcal{A}$ is spanned by $A$ and $B$ under the same set of operations. States for this algebra may be taken to be all probability density distributions $p(x)\geq 0$, $\int_{0}^{3} p(x) dx =1$. The expectation value for an observable $f$ with respect to a state $p$ is then given by $\langle f \rangle_p= \int_{0}^{3} f(x) p(x) dx = \sum_i p_i f_i $, with $p_i \equiv \int_{0}^{3} \chi_i p(x) dx$. It is clear that: (i) two algebra elements (observables) $f$ and $g$ are ordered as $f\geq g$, i.e., $\langle f \rangle_p\geq \langle g \rangle_p$  for all $p$, if and only if $f_i\geq g_i$ for all $i=1,2,3$. Thus, three experimentally accessible ``peaked enough'' states such as
\[
p_i(x)=\begin{cases}
      2 & x\in \left[i-1+\frac{1}{4},i-\frac{1}{4}\right]\\
	0	& \textrm{elsewhere}
     \end{cases},
\]
are sufficient to determine whether any two observables in the algebra $\mathcal{A}$ are ordered.
In particular, by checking said states one finds that the two observables $A$ and $B$ are actually not ordered, therefore they cannot be used to test any quantumness in the sense of Theorem 1.

\begin{figure}[!t]
\begin{center}
\includegraphics[width=0.35\textwidth]{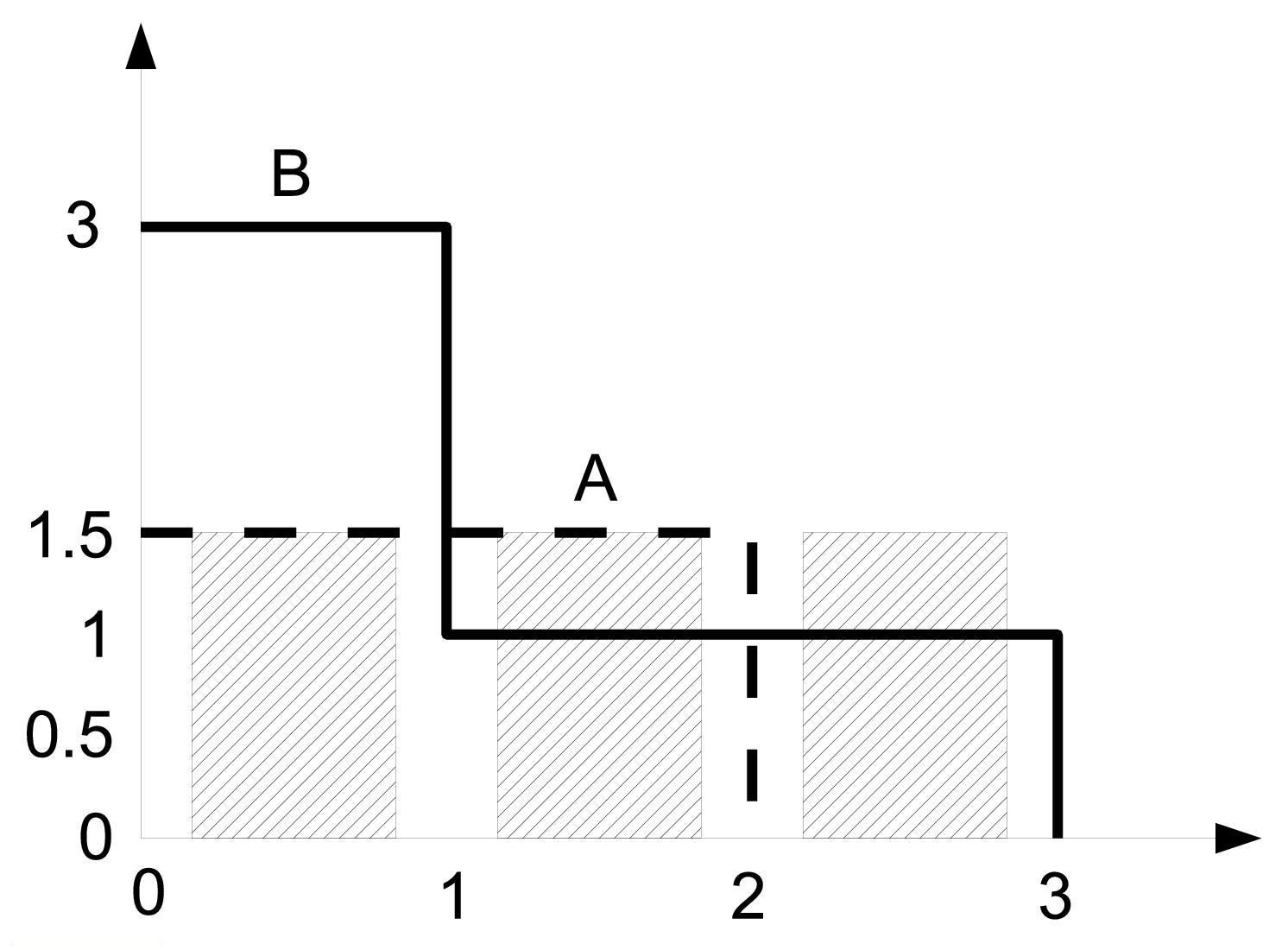}
\caption{The absence of ordering between the observables $A$ and $B$ is revealed by set of experimentally accessible states. Observable functions are depicted as segment lines. Hatched rectangles correspond to states.}
\label{fig:algebra}
\end{center}
\end{figure}

\begin{figure}[!t]
\begin{center}
\includegraphics[width=0.35\textwidth]{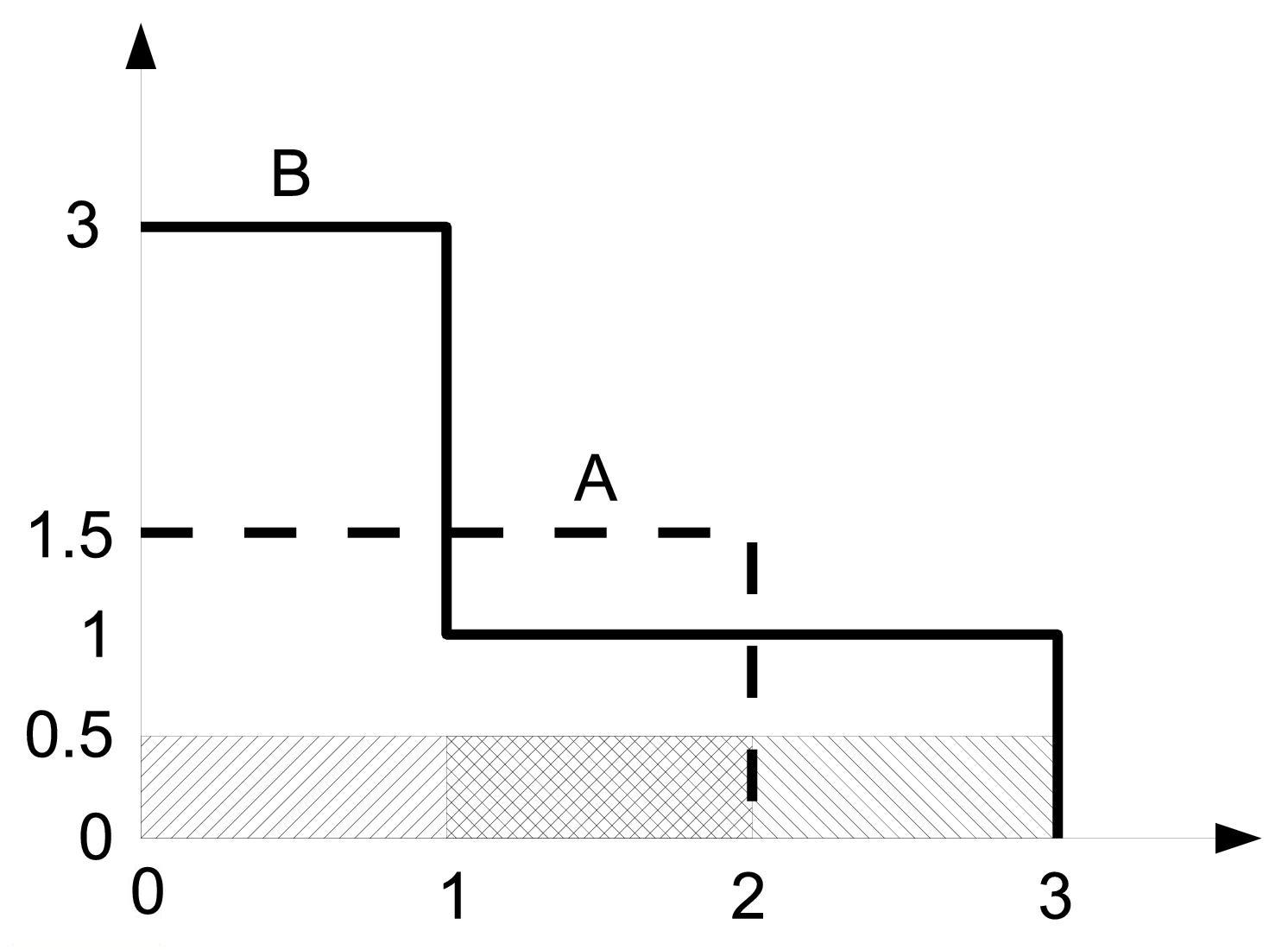}
\caption{Minimality loophole: experimental states that are too coarse-grained may erroneously lead to claims of quantumness of an experiment which admits a non-minimal classical algebraic model. Observable functions are depicted as segment lines. Hatched rectangles correspond to states.}
\label{fig:loophole}
\end{center}
\end{figure}

\noindent {\bf Example 3}
Consider the same setting as in Example 2, in particular the same observable functions $A$ and $B$. However, let us consider more coarse grained experimentally accessible states (see Figure~\ref{fig:loophole})
\[
q_1(x)=\begin{cases}
      \frac{1}{2} & x\in [0,2]\\
	0	& \textrm{elsewhere}
     \end{cases},
\quad
q_2(x)=\begin{cases}
      \frac{1}{2} & x\in [1,3]\\
	0	& \textrm{elsewhere}
     \end{cases}.
\]
The latter states are such that
\[
\langle B \rangle_{q_i}>\langle A \rangle_{q_i}>0 \quad \textrm{for}\quad i=1,2,
\]
but while $\langle B^2 \rangle_{q_1}> \langle A^2 \rangle_{q_1}>0$, one has
$\langle B^2 \rangle_{q_2}-\langle A^2 \rangle_{q_2}=-0.125<0$. Thus, it is clear that the choice of observables and states for the experiment does not admit a minimal classical model. While one could be led to conclude that the interpretation of the experiment requires a quantum model, it is clear that the model is classical by construction. In any case, said classical model is non-minimal, i.e. ordering on the experimental states does not lead to ordering for the full model. This is due to the fact that the experimentally accessible states do not have a high enough ``resolution''.

\subsection{ A single qubit}\label{sec:singlequbit}
As an example we would like to find the optimal quantumness witness $V= B^2-A^2$ for a qubit
which has a maximal magnitude of its negative eigenvalue under a certain normalization condition. Such an optimal choice should be useful for designing experiments with particular implementations of qubits \cite{Gen}.
In contrast to the previous paper \cite{AV} where a numerical solution was found for the
normalization condition $0\leq A\leq B\leq\openone$, here we find an analytical solution for the more convenient normalisation
\begin{equation}
\mathrm{Tr}\{B\}=2.
\label{tr}
\end{equation}
We begin with the following parameterisation in terms of Pauli matrices $\vec{\sigma} = (\sigma_1 ,\sigma_2 ,\sigma_3)$:
\begin{equation}
A=a_0\openone+\vec{a}\cdot\vec{\sigma}\mbox{ and }
B=\openone+\vec{b}\cdot\vec{\sigma},\label{eqn:defAB}
\end{equation}
and introduce the following scalar parameters:$p=|\vec{a}|^2$, $q=|\vec{b}|^2$, $r=|\vec{b}-\vec{a}|^2$ and $s=a_0$. The condition $0\leq A\leq B$ can be expressed
by the inequalities
\begin{equation}
s^2 \geq p\ \mathrm{ and }\ (1-s)^2\geq r.
\label{ine}
\end{equation}
The lower eigenvalue of $V= B^2-A^2$ is calculated to be
\begin{equation}
\lambda_V^-=1-p+q-s^2-2 \sqrt{q-q s+(r+p (-1+s)) s}.\label{eqn:lef}
\end{equation}
In minimising $\lambda_V^-$, it can easily be seen that  $r$ should be taken to be a maximum, thereby making the second inequality in equation (\ref{ine}) an equality. This equality means that for the lower eigenvalue of $M$ to be a minimum, the determinant of $B-A$ must be zero. Additionally, one of the eigenvalues of $B-A$ is zero, and the other positive. This satisfies the ordering condition $0\leq A\leq B$. This naturally leads us to a different, more convenient, parameterisation. We now choose $A$ and $B$ such that
\begin{equation}
A=\left(
\begin{array}{cc}
 -2 (-1+t+u) & 2 z \\
 2 z & 2 u
\end{array}
\right)\mbox{ and }
B=\left(
\begin{array}{cc}
 2-2 u & 2 z \\
 2 z & 2 u
\end{array}
\right).
\end{equation}
This immediately gives us
\begin{equation}
B-A=\left(
\begin{array}{cc}
 2 t & 0 \\
 0 & 0
\end{array}
\right)
\end{equation}
and
\begin{equation}
V=B^2-A^2=\left(
\begin{array}{cc}
 -4 t (-2+t+2 u) & 4 t z \\
 4 t z & 0
\end{array}
\right).
\end{equation}
The lower eigenvalue of $V$ is now calculated to be
\begin{equation}
\lambda_V^-=-2 \left(t (-2+t+2 u)+\sqrt{t^2 \left((-2+t+2 u)^2+4 z^2\right)}\right).\label{eqn:lnew}
\end{equation}
The inequality in Eq. (\ref{ine}) expressed in terms of these new parameters can be written as
\begin{equation}
z^2\leq u(1-u-t).
\label{ine2}
\end{equation}
It is easy to see that in order to minimise Eq. (\ref{eqn:lnew}), the parameter $z$ should be a maximum under the constraints. This results in Eq. (\ref{ine2}) being an equality, giving us the lower eigenvalue of the witness $V$, written in terms of two parameters $u$ and $t$, as
\begin{equation}
\lambda_V^-=-2 \left(\sqrt{t^2 \left((-2+t)^2-4 u\right)}+t (-2+t+2 u)\right).
\end{equation}
To solve this equation, the partial derivatives must be zero at the extrema of the lower eigenvalue. The partial derivatives are
\begin{equation}
\frac{\partial \lambda_V^-}{\partial u}=4 t \left(-1+\frac{1}{\sqrt{(t-2)^2-4 u}}\right)\label{eqn:par1}
\end{equation}
and
\begin{equation}
\frac{\partial \lambda_V^-}{\partial t}=-4 \left(-1+t+\frac{(2+t(t-3)-2 u)}{\sqrt{(-2+t)^2-4 u}}+u\right).\label{eqn:par2}
\end{equation}
In solving $\frac{\partial \lambda_V^-}{\partial u}=0$, and using the fact that $t$ is nonzero, we can write
\begin{equation}
(t-2)^2=4u+1.
\end{equation}
This is the same as writing
\begin{equation}
\det{A}=0,\label{eqn:detAzero}
\end{equation}
or equivalently that the lower eigenvalue of $A$ be zero. Substituting $u=\frac{(t-2)^2-1}{4}$ into Equation (\ref{eqn:par2}) we can then write
\begin{equation}
\frac{\partial \lambda_V^-}{\partial t}=-\frac{(t-1) \left(3t^2-t\right)}{t}=(t-1) \left(3t-1\right).
\end{equation}
We do not wish to take the trivial solution $t=1$, and this leaves us with $$u=\frac{4}{9} \mbox{ and } t=\frac{1}{3}.$$ The minimum is now found to be
\begin{equation}
\lambda_{V(min)}^-=-\frac{4}{27}.
\end{equation}
It is convenient to use a parameterisation with one of the operators (say $B$) being diagonal.
A straightforward computation leads to
\begin{equation}
A=\left(
\begin{array}{cc}
 \frac{2}{99} \left(33+5 \sqrt{33}\right) & \frac{4 \sqrt{\frac{2}{33}}}{3} \\
 \frac{4 \sqrt{\frac{2}{33}}}{3} & \frac{2}{3}-\frac{10}{3 \sqrt{33}}
\end{array}
\right)\mbox{ and }
B=\left(
\begin{array}{cc}
 \frac{1}{9} \left(9+\sqrt{33}\right) & 0 \\
 0 & \frac{1}{9} \left(9-\sqrt{33}\right)
\end{array}
\right).\label{eqn:maxviolation}
\end{equation}
with the eigenvector corresponding to $\lambda_{V(min)}^-=-\frac{4}{27}$
given by $(\frac{2}{3}\sqrt{2},\frac{1}{3})$. \\

\begin{figure}[htp]
  \begin{center}
    \subfigure[Expectation value of $B-A$.]{\label{fig:Vmap}\includegraphics[scale=0.4]{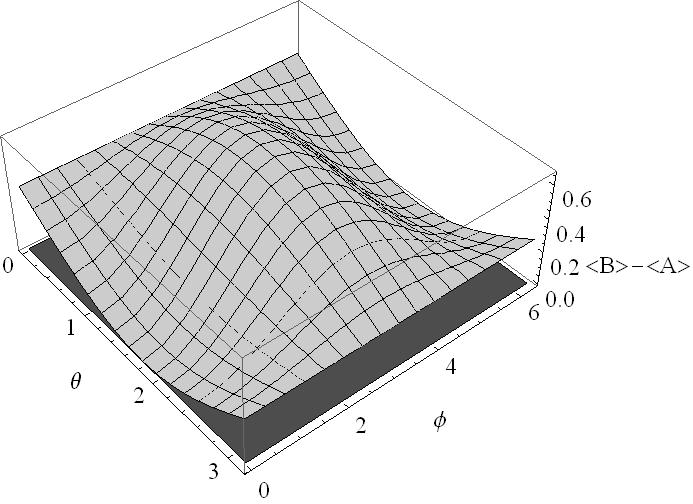}}\ \ \ \ \ \ \ \
    \subfigure[Expectation value of $B^2-A^2$.]{\label{fig:Wmap}\includegraphics[scale=0.4]{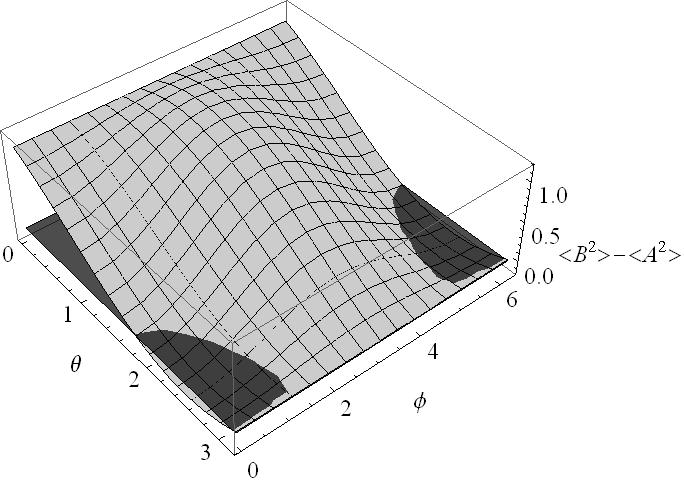}} \\
  \end{center}
  \caption{Expectation values of $B-A$ and $B^2-A^2$ on the pure state $\psi(\theta,\phi)$.}
  \label{fig:VWmaps}
\end{figure}

The parameters for the observables $A$ and $B$ in Equation (\ref{eqn:maxviolation}) lead to a maximal violation of the ordering condition $0\leq A^2\leq B^2$. Figure \ref{fig:VWmaps} may be helpful for designing an experiment for a single qubit to test for such a violation.

Figure \ref{fig:VWmaps} shows the mean values of the operators  $\langle B-A\rangle$ and $\langle B^2-A^2\rangle$ on the pure states of a single qubit
\begin{equation}
\ket{\psi(\theta,\phi)}=\cos(\frac{\theta}{2})\ket{0}+\exp(i\phi)\sin(\frac{\theta}{2})\ket{1},
\end{equation}
where $0\leq\theta\leq\pi$ and $0\leq\phi\leq 2\pi$. In both pictures the $x$ and $y$ axes are the parameters $\theta$ and $\phi$ respectively, while the $z$-axis is the range of mean values of $B-A$ and $B^2-A^2$. The dark gray horizontal plane is given by $z=0$.\\

In Figure \ref{fig:Vmap}, it is clear that the expectation $\langle B-A\rangle$ is never negative. This satisfies the ordering condition from Equation (\ref{order}), $A\leq B.$\\

The behaviour of the expectation $\langle B^2-A^2\rangle$ is much more interesting, and is shown in Figure \ref{fig:Wmap}. In this picture, one can clearly see that $\langle B^2-A^2\rangle$ becomes negative around $\theta=\frac{2}{3}\pi$ and $\phi=0$. This area is a connected subset of the Bloch sphere and should be the focus of any experiment aiming to determine whether or not the behaviour of a system may be described by classical theories.

\subsection{Additive observables for N-body systems}

Very often the set of accessible observables is rather restricted and includes only the observables with a particular structure. For example, for many-body systems we usually restrict ourselves to collective, global, observables $A$ and $B$ of the form
\begin{equation}
A=\frac{1}{N}\sum_{i=1}^Na_i\ ,\
B=\frac{1}{N}\sum_{i=1}^Nb_i
\label{collective}
\end{equation}
where $a_i$ and $b_i$ are copies of the single particle observables $a$ , $b$ and $N$ is the number of particles. Assume that $0\leq a\leq b$ and the single particle quantumnes witness $v= b^2-a^2$ has minimal negative eigenvalue $-\mu$.
With these collective observables, it is interesting to study their behavior as $N$ becomes large. Intuitively one can expect that as $N$ increases, so any detectable quantum behavior should diminish. This can be seen for the collective witness $V=B^2-A^2$ which is rewritten in terms of "diagonal" and "off-diagonal" terms as
\begin{equation}
V=\frac{1}{N^2}\sum_{i=1}^N(b_i^2-a_i^2)+\frac{1}{N^2}\sum_{i=1}^N\sum_{j=1,i\neq j}^N(b_ib_j-a_ia_j).
\label{colobs}
\end{equation}
The first sum is simply the sum of the individual witnesses, $v_i$, and the second term is  positive. This second sum can be rewritten as
\begin{equation}
\frac{1}{N^2}\sum_{i=1}^N\sum_{j=1,i\neq j}^N(b_ib_j-a_ia_j)=\frac{1}{N^2}\sum_{i=1}^N\sum_{j=1,i\neq j}^N(b_i+a_i)(b_j-a_j),
\end{equation}
where the terms $a_ib_j-b_ia_j$ disappear under the sum. Since we have the condition $b_i\ge a_i\ge 0$, and since the operators  $a_i,b_i$ and $a_j,b_j$ act on different spaces for $i\neq j$, it can be seen that this sum is positive.
Therefore, the magnitude of the lowest negative eigenvalue of $V$ is bounded from above by $\mu/N$ and quantum behaviour becomes undetectable for collective observables and large $N$.\\
This should have consequences for the discussion of the quantumness of superconducting qubits and BEC. Namely, for both cases the accessible observables are particle numbers or electric charge in a macroscopic region, macroscopic current, etc., which are all of the collective type (\ref{collective}) with the typical values of  $N=10^5$ to $10^9$.

\section{Bell inequalities and quantumness witness}

Bell inequalities, in various formulations, characterize correlations between measurement outcomes which admit the so-called Local Hidden Variable Model (LHVM). Here, according to the discussion in Section 2.3 we  replace the LHVM by a CAM.
For simplicity we consider only the Clauser, Horne, Shimony and Holt version (CHSH inequality) where the observables $A_1, A_2$ are measured on the one subsystem and the observables $B_1, B_2$ on the other one. We assume that all observables are dichotomic, i.e. their outcomes are equal to $\pm 1$. As discussed in Section 2.3, in this  model all observables are treated as classical functions
$A_i(\lambda) , B_j(\lambda)$ on a common space $\Lambda$ and take the values $\pm 1$ only. Then the observables
\begin{equation}
X = 2 \pm(A_1 B_1 + A_1 B_2) \ , Y= 2 \pm(A_2B_1 - A_2B_2) .
\label{bellob}
\end{equation}
are positive and hence $XY\geq 0$ as well. Therefore, for any probability distribution $P(\lambda)$ we obtain the inequality
\begin{equation}
0\leq \frac{1}{2}\langle X Y\rangle_P = 2 \pm\langle(A_1 B_1 + A_1 B_2 + A_2B_1 - A_2B_2\rangle_P,
\label{bell1}
\end{equation}
which is equivalent to the standard form of CHSH inequality
\begin{equation}
|\langle(A_1 B_1 + A_1 B_2 + A_2B_1 - A_2B_2)\rangle_P| \leq 2\ .
\label{bell}
\end{equation}
The experimentally observed  violations of the inequality (\ref{bell}) means that Nature cannot be described by a CAM.\\

\subsection {Test of quantumness for a bipartite system}

The CHSH inequality (\ref{bell}) can also be used to test whether a quantum device  does not operate in the  semiclassical regime. It seems to be even more reliable than the single-system test discussed in Section 4. Namely, we now do not have the \emph{minimality loophole} problem because any violation of (\ref{bell}) means that the description in terms of CAM is not valid.\\
However, to perform the test of (\ref{bell}) we need, firstly, to control precisely the system composed of two subsystems which is usually much more difficult than in the case of a single system. Secondly, we have to perform simultaneously two independent measurements on both subsystems. Even if the system is essentially classical it is often very difficult to satisfy
the independence condition (\ref{ind}) which leads to the so-called \emph{locality loophole}. Therefore, for many implementations of quantum devices relevant for quantum information processing the single-system test might be much easier to perform than the bipartite one.

\subsection{Bell observables and quantumness witness}
 Bearing in mind the definitions (\ref{bellob}), the relation to quantumness witnesses is rather obvious. Consider now the general case of four dichotomic quantum observables $A_1, A_2, B_1, B_2$. Replacing the product of functions by a symmetrized product $A_i\circ B_j=\frac{1}{2}\{A_i,B_j\}$ we obtain the positive observables $X$ and $Y$ for a general quantum
model, and $C= XY + YX$ is a  quantumness witness. However,  neither the observable $C$ nor its mean value is operationally
well-defined. Let us assume that $\{A_i\}$ and $\{B_j\}$ correspond to different subsystems, $[A_i,B_j]=0$ and the mean value of the following \emph {Bell observable }${\cal B}$ (compare with \cite{Bra}) has an operational meaning as a combination of measurable correlations
\begin{equation}
{\cal B} = 2\pm (A_1 B_1 + A_1 B_2 + A_2 B_1 - A_2 B_2)\ .
\label{bellobs}
\end{equation}
In this case the Bell observable is related to the quantumness witness $C$ by the expression
\begin{equation}
2{\cal B} =  C + [A_1 , A_2][ B_1 , B_2]\ ,
\label{bellwit}
\end{equation}
which provides an interesting relation between Bell inequality, QW and the Heisenberg uncertainty relations (compare
with \cite{Ter} for Bell inequalities vs entanglement witnesses).

\section{Generalizations of QW}
The presented examples of quantumness witnesses suggest a certain generalization of this idea. Take two observables described by
self-adjoint operators $R$ and $S$ with the spectra ${\rm Spec}(R)$ and ${\rm Spec}(S)$ respectively.
\par
An observable $W= W(R;S)$ written as an (ordered) polynomial or power series of $R ,S$ is a QW if  it possesses at least one negative eigenvalue and the function $w(r;s)$ obtained by the replacement of operators by real numbers, $R\to r$ , $S\to s$ in $W(R;S)$ is positive, i.e.
$w(r;s)\geq 0$ for all $r\in {\rm Spec}(R)$ and $s\in{\rm Spec}(S)$.
\par
The previously introduced QW of the form $V = B^2 - A^2$  satisfies this definition taking $R\equiv A$ and $B \equiv R + S$ with $R\geq 0$ and $S\geq 0$.
Indeed, $V\equiv W(R;S)= R^2 +RS +SR$ possesses a negative eigenvalue and $w(r;s) =r^2 +2rs \geq 0$ for $r,s\geq 0$. The same holds for QW of the type $C= XY + YX$ with the identification $X=R\geq 0$, $Y= S\geq 0$.
\par
Another class of quantumness witnesses is related to phase-representations of Quantum Mechanics in terms of coherent states and the so-called \emph{ P-representation} \cite{Per,Kla}. Consider a single quantum oscillator with the canonical pair of observables $Q, P$, the  annihilation operator
$a = \frac{1}{\sqrt{2}}(Q-iP)$ satisfying
\begin{equation}
[Q , P] =  i\ , [a , a^{\dagger}]= 1
\label{qp}
\end{equation}
and the family of coherent vectors
\begin{equation}
a |z\rangle = z |z\rangle , z=\frac{1}{\sqrt{2}}(q-ip)\ , q,p\in \textbf{R}
\label{coh}
\end{equation}
with the normalisation and completeness conditions
\begin{equation}
\langle z|z\rangle = 1\ ,\ \int d^2\,  |z\rangle\langle z| =  I\ .
\label{coh1}
\end{equation}
Any density matrix $\rho$  can be represented by a function (distribution) on the phase-space denoted by $\tilde{\rho}(z)$ and
such that (P-representation)
\begin{equation}
\int d^2 z\, \tilde{\rho}(z)|z\rangle\langle z| =  \rho\ .
\label{PQ}
\end{equation}
The \emph{dequantisation} map for the states $\rho \to{\tilde\rho}(z)\equiv{\tilde\rho}(q;p)$ is linear but
does not preserve positivity. Therefore, the function ${\tilde\rho}(z)$ is called quasi-probability distribution and the states which have positive ${\tilde\rho}(\alpha)$
can be considered as  "classical" at least with respect to this particular phase-space representation.
\par
We now construct a whole family of QWs in the form of a series
\begin{equation}
W(Q;P) = \sum_{m,n} c_{mn} \bigl(a^{\dagger}\bigr)^m a^n
\label{qwser}
\end{equation}
such that the function
\begin{equation}
w(q;p) = \sum_{m,n} c_{mn} {\bar z}^m z^n\geq 0
\label{qwser1}
\end{equation}
and $W(Q;P)$ possesses at least one negative eigenvalue. Obviously, for any classical state $\rho$
\begin{equation}
{\rm Tr}(\rho W) = \int dqdp \, w(q;p) {\tilde\rho}(q; p))\geq 0
\label{ave}
\end{equation}
and hence $W(Q;P)$ detects certain "non-classical" states.

A simple example is provided as the observable $K_m$  with $m\geq 1$ defined  by
\begin{equation}
K_m = (a^{\dagger})^2  a^2 - 2m  a^{\dagger}a +m^2 = N^2 - 2(m+\frac{1}{2}) N + m^2
\label{psW}
\end{equation}
with $N= a^{\dagger}a$, $k_m(z) = (|z|^2 -m)^2$. It is a phase-space QW because $K_m$ has negative eigenvalues corresponding to the eigenvectors $|n>$ of $N$ with $n \in (m+\frac{1}{2}- \sqrt{m +\frac{1}{4}}, m+\frac{1}{2}+ \sqrt{m +\frac{1}{4}})$.\\

\section{Conclusions}
Motivated by the idea of entanglement witnesses we have introduced special types of quantum observables called quantumness witnesses (QW). The main feature of a QW is  that for any state which is considered in a given context as  "classical" its mean value is positive but nevertheless its spectrum contains at least one negative eigenvalue. The notion
of classicality is understood here in a very practical and "contextual" sense. Namely, we assume that we have at our disposal
a restricted set of experimental data which we want to interpret in terms of sets of accessible states and observables within a classical or quantum model. A properly chosen QW can provide an experimental test which detects among the accessible states those which cannot be described by a classical model. The necessary condition is the possibility of computing the average of the QW using experimental data. A natural application of this idea is the implementation of quantum information processing. Devices designed for quantum information processing must work in the quantum regime and therefore violation of classicality described in terms of QWs could be the first experimental test of their usefulness.

\emph{Acknowledgements.} The authors thank  M. Horodecki, R. Horodecki, and M. Genovese  for
discussions.   Financial support by the POLAND/SA COLLABORATION PROGRAMME of the National Research Foundation of South Africa and the Polish Ministry of Science and Higher Education and by the
European Union through the Integrated Project SCALA is acknowledged.



\begin{thebibliography}{99}
\bibitem{AV}
R. Alicki and N. Van Ryn, J. Phys. A: Math. Theor. {\bf 41} 062001 (2008)
\bibitem{dec}
E. Joos, H.D. Zeh, C. Kiefer, D. Giulini, J. Kupsch, and I,-O. Stamatescu, \emph{Decoherence and the Appearance of a Classical World in Quantum Theory}, 2nd ed.
Springer,Berlin 2003.
\bibitem{NBS}
Y. Nakamura et.al., Nature {\bf 398} (1999), 786;
A.J. Berkley et. al., Science {\bf 300} (2003), 1548;
M. Steffen et.al., Science {\bf 313}, 1423, (2006).
\bibitem{BEC}
N. Teichmann and  C. Weiss, Europhys.Lett.{\bf 78}, 10009, (2007).
\bibitem{Ahn}
J. Ahn, T.C. Weinacht and P.H. Bucksbaum, Science {\bf 287}, 463 (2000).
\bibitem{Gro} N. Gr\o nbech-Jensen, et.al. Phys.Rev.Lett.{\bf 93}, 107002, (2004);
N. Gr\o nbech-Jensen and M. Cirillo,  Phys.Rev.Lett.{\bf 95}, 067001, (2005);
J.E. Marchese, M. Cirillo and N. Gr\o nbech-Jensen, cond-mat/0604111, (2006).
\bibitem{ali}
R. Alicki, arXiv:quant-ph/0610008
\bibitem{Gen}
 G.Brida, I. Degiovanni, M. Genovese, V. Schettini, S. Polyakov, and A. Migdall,
``Experimental test of nonclassicality for a single particle'', arXiv:0804.1646.
\bibitem{Alb}
G. Alber , T. Beth, M. Horodecki,P. Horodecki,R. Horodecki, M. Roetteler, H. Weinfurter, R. Werner and A. Zeilinger, \emph{Quantum Information. An Introduction to Basic Theoretical Concepts and Experiments},
STMP 173, Springer , Berlin, 2001
\bibitem{Dix}
J. Dixmier, \emph {C*-algebras}, (translated from the French by Francis Jellett)  North-Holland , Amsterdam, 1977.
\bibitem{Kad}
R.V. Kadison and J.R. Ringrose,\emph{ Fundamentals of the Theory of Operator Algebras: Elementary Theory}, Academic Press, New York ,1983
\bibitem{Oga}
T. Ogasawara, J.Sc. Hiroshima Univ.{\bf 18}, 307, (1955)
\bibitem{Dick}
W.M. Dickson, \emph{Quantum chance and non-locality}, Cambridge University Press, Cambridge, 1998.
\bibitem{Ind}
This condition is called "$\lambda$- dependence " in \cite{Dick}.
\bibitem{H4}
R. Horodecki, P. Horodecki, M. Horodecki, and K. Horodecki, arXiv:quant-ph/0702225
\bibitem{bell}
J.S. Bell,  Rev. Mod. Phys. \emph{38}, 447 (1966)
\bibitem{Bra}
S. Braunstein, A. Mann and M. Revzen, Phys.Rev.Lett.{\bf 68}, 3259, (1992)
\bibitem{Ter}
B.M. Terhal, Phys.Lett.{\bf A271}, 319, (2000).
\bibitem{Per}
A. Perelomov, \emph{Generalized Coherent States and Their Applications}, Springer, Berlin, 1986.
\bibitem{Kla}
J.R. Klauder and B.-S. Skagerstam, \emph{Coherent States. Applications in Physics and Mathematical Physics}, World Scientific,
Singapore, 1985.
\bibitem{Car}
H.J. Carmichael, \emph{Statistical Methods in Quantum Optics 2}, Springer, Berlin, 2008.
\bibitem{W}
R. F. Werner, Phys. Rev. A {\bf 40}, 4277, 1989.
\end{thebibliography}
\end{document}